\documentclass[prl,superscriptaddress,twocolumn,showpacs,preprintnumbers,%
amsmath,amssymb]{revtex4-1}

\usepackage{graphicx}
\usepackage{dcolumn}
\usepackage{bm}

\begin{document}   
\title{Self-energy effects in electronic Raman spectra of
doped cuprates due to magnetic fluctuations}
\author{Roland Zeyher}
\affiliation{Max-Planck-Institut f\"ur Festk\"orperforschung,
             Heisenbergstrasse 1, D-70569 Stuttgart, Germany}

\author{Andr\'{e}s Greco}
\affiliation{Departamento de F\'{\i}sica, Facultad de Ciencias Exactas
e Ingenier\'{\i}a and IFIR(UNR-CONICET), 2000 Rosario, Argentina}

\date{\today}

\begin{abstract}
We present results for magnetic excitations in doped copper oxides using
the random phase approximation and itinerant electrons. 
In the [1,0] direction the observed excitations
resemble dispersive quasi-particles both in the normal and superconducting state
similar as in recent resonant inelastic X-ray scattering (RIXS)
experiments. In the [1,1] direction the excitations
form, except for the critical region near the antiferromagnetic wave
vector ${\bf Q}=(\pi,\pi)$, only very broad continua. Using the obtained spin
propagators we calculate electron self-energies and their effects on
electronic Raman spectra. We show that the recently observed additional 
peak at about twice the pair breaking in  B$_{1g}$ symmetry below T$_c$ in
HgBa$_2$CuO$_{4+\delta}$ can be explained as a self-energy effect
where a broken Cooper pair and a magnetic excitation appear as final states. The
absence of this peak in B$_{2g}$ symmetry, which probes mainly electrons near
the nodal direction, is explained by their small self-energies compared to those
in the antinodal direction.  
\end{abstract}


\maketitle

Electronic Raman spectra in doped cuprates are dominated by a pair 
breaking peak in the superconducting state whereas in the normal state they are 
rather structureless. The main
features of the spectra in the superconducting state can be explained 
within BCS theory using free
quasi-particles. \cite{Devereaux} 
On the other hand RIXS experiments have shown the existence of spin 
fluctuations in doped cuprates
in form of dispersive, broad but well-defined excitations, 
both in the 
normal and the superconducting state. \cite{LeTacon,Dean} The interaction 
between spin fluctuations and
electrons gives rise to self-energies which have been used to explain,
for instance, the observed kinks in the electronic dispersion in many high-T$_c$
superconductors. \cite{Zeyher2,Dahm,Manske}
Self-energy effects should also be present in electronic Raman
spectra causing for instance the recently observed broad peak above 
the pair breaking peak in HgBa$_2$CuO$_{4+\delta}$. \cite{Li}
It is the purpose of this Letter to calculate the properties of spin 
fluctuations, the resulting self-energies and their effects on Raman scattering
and to compare the results with recent experiments. \cite{LeTacon,Li}
Our calculations are 
based on an itinerant picture of magnetism and    
will be limited to the optimally doped and overdoped region
to avoid unsolved problems with the existence and nature of the pseudogap
in the underdoped regime.

The Raman susceptibility can be written in the superconducting 
state as, \cite{Zeyher1}
\begin{eqnarray}
\chi_\alpha(i\nu_m) = \frac{1}{N_c} \sum_{\bf k} \gamma_\alpha^2({\bf k})(
\Pi_{11,11}({\bf k},i\nu_m) + \Pi_{22,22}({\bf k},i\nu_m) \nonumber \\
-\Pi_{12,21}({\bf k},i\nu_m)-\Pi_{21,12}({\bf k},i\nu_m)), \hspace{0.5cm}
\label{chi}
\end{eqnarray}
\begin{equation}
\Pi_{ij,kl}({\bf k},i\nu_m) = \int_{-\infty}^{\infty} d \epsilon d \epsilon' 
A_{ij}({\bf k},\epsilon) A_{kl}({\bf k},\epsilon') 
\frac{f(\epsilon')-f(\epsilon)}{i\nu_m+\epsilon' -\epsilon}.
\label{P}
\end{equation}
The indices $i,j,$... assume the values 1 or 2, $\nu_m$ denotes 
a bosonic Matsubara
frequency, $f$ the Fermi function, and $N_c$ is the number of primitive cells.
Assuming a square lattice and using the lattice constant as the length unit
$\gamma_\alpha({\bf k})$ stands for the form factors 
$\gamma_1({\bf k})= t(\cos k_x + \cos k_y) +4t'\cos k_x \cos k_y$, 
$\gamma_3({\bf k}) = t(\cos (k_x-\cos k_y)$,
$\gamma_4({\bf k})= -4t'\sin k_x \sin k_y$, corresponding to the representations
A$_{1g}$, B$_{1g}$, and B$_{2g}$ of the point group D$_{4h}$,
respectively. $t$ and $t'$ are effective nearest and second-nearest neighbor 
hopping amplitudes. 

Using the Nambu representation the inverse of the $2\times 2$ 
electron Green's function matrix $G(k,i\omega_n)$ is
\begin{eqnarray}
G^{-1}({\bf k},i\omega_n) = (i\omega_n-\Sigma_+({\bf k},i\omega_n))\tau_0 
\nonumber \\
-(\epsilon({\bf k})+ \Sigma_-({\bf k},i\omega_n)) \tau_3 
- \Delta({\bf k}) \tau_1.
\label{G-1}
\end{eqnarray}
$\tau_1$ and $\tau_3$ are Pauli matrices, $\tau_0$ the $2\times 2$ unit matrix, and $\omega_n$
a fermionic Matsubara frequency. 
$\Sigma_{\pm}({\bf k},i\omega_n)$  stands for 
$(\Sigma_{11}({\bf k},i\omega_n) \mp \Sigma_{11}(-{\bf k},-i\omega_n))/2$, 
where $\Sigma_{11}({\bf k},i\omega_n)$
is the (1,1) component of the 2x2 self-energy matrix. 
$\epsilon({\bf k})$ is the electron energy
\begin{equation}
\epsilon({\bf k}) = -2t(\cos k_x + \cos k_y) - 4t'\cos k_x \cos k_y -\mu,
\label{epsilon}
\end{equation}
and the spectral function $A_{ij}$ is defined by 
$-1/\pi Im G_{ij}({\bf k},\omega+i\eta)$.
The self-energy $\Sigma$ is calculated from the Fock diagram
with a bosonic propagator describing spin fluctuations. To simplify the
calculation we do not consider the non-diagonal part of $\Sigma$ assuming that
it has already been taken into account by the phenomenological 
gap parameter $\Delta({\bf k})$. Assuming also that the frequency-independent
Fock term has been accounted for in the phenomenological hoppings
$t$ and $t'$ the frequency-dependent
part of $\Sigma_{11}$ is given by \cite{Zeyher2} 
\begin{eqnarray}
\Sigma_{11}({\bf k}, i\omega_n) = \nonumber \\
-\frac{T}{2N_c} \sum_{{\bf q},m} J^2({\bf q}) 
\chi({\bf q},i\nu_m) G_{11}^{(0)}({\bf k}+{\bf q},i\omega_n+i\nu_m).
\label{sigma}
\end{eqnarray}
$\chi({\bf k},i\nu_m)$ is the zz component of the spin susceptibility which 
assumes in the RPA the form \cite{Zeyher2, Eschrig} 
\begin{equation}
\chi({\bf k},i\nu_m) = \frac{\chi^{(0)}({\bf k},i\nu_m)}{1-J({\bf k})\chi^{(0)}({\bf k},i\nu_m)}.
\label{spin}
\end{equation}
$\chi^{(0)}({\bf q},i\nu_m)$ represents the bare spin bubble. 
$J({\bf k})$ is equal to 
$2J(\cos k_x + \cos k_y)$ where $J$ is the Heisenberg constant. 
Since self-energy effects in Raman spectra are expected to be rather small we
may use $G^{(0)}_{11}$ instead of $G_{11}$ in Eq.(\ref{sigma}).

In the following calculations we used the parameters $t'/t=-0.25$ and  
$\mu/t=-0.9$ corresponding to the doping $\delta = 0.15$, 
where $t$
is the effective nearest-neighbor hopping amplitude and about $0.15 eV$ 
in optimally doped YBCO. \cite{Norman}  
We expect a similar value in the case of  HgBa$_2$CuO$_{4+\delta}$.
$J$ was determined so that the experimental value of 0.056 eV \cite{Yu}
for the magnetic resonance energy in the 
superconducting state was reproduced yielding $J/t=1.30$.
The bare order parameter $\Delta$ was fixed to $\Delta/t=0.45$ which reproduces
roughly the position of the observed pair breaking peak in B$_{1g}$ 
symmetry in the sample with the highest T$_c$ of 94K of Ref. \cite{Li}. 
 
An important ingredient for the calculation of $\Sigma$ is the 
dynamical spin susceptibility
$\chi({\bf k},\omega)$. Its imaginary part $\chi''({\bf k},\omega)$ 
has been discussed extensively in
the literature \cite{Eschrig} in conjunction with the resonant mode, 
i.e., in the 
superconducting state
and near the antiferromagnetic wave vector ${\bf Q}$. We extend these 
calculations to
wave vectors throughout the Brillouin zone (BZ) and also to the normal state. 
Such calculations
are interesting in view of recent RIXS measurements \cite{LeTacon} which 
cover a more extended region.
The upper and lower diagrams in Fig. 1 show $\chi''({\bf k},\omega)$ 
along the $[1,1]$ direction in the normal and
the superconducting state, respectively. For wave vectors away 
from ${\bf Q}$ (and also from the origin (0,0))
the spectra are extremely broad and structureless, exhibiting often 
several flat maxima. Only
the center of gravity of the lines indicates an increase with momentum 
up to around the
middle of the zone and then a pronounced decrease towards ${\bf Q}$. 
In the normal state
the curve for momentum ${\bf Q}$ shows a well-pronounced critical peak 
due to the proximity of the
antiferromagnetic instability. In the superconducting state this peak 
moves to somewhat higher
energies and becomes the resonant peak with a finite width due to our 
employed damping constant 
$\eta = 0.1$. 
Being a bound state inside the gap it attracts a huge 
amount of spectral weight. This holds to a lesser degree even for the 
momentum 3$\bf Q$/4 where
the structureless continuum in the normal state transforms into a 
well-pronounced peak in the presence of superconductivity.

$\chi''({\bf k},\omega)$ looks quite different in the $[1,0]$ direction 
as illustrated in Fig. 2. It shows both 
in the normal and the superconducting state well-defined and narrow peaks which disperse
as a function of momentum similar as the spin waves of a Heisenberg model.
However, in our case the sharp peaks do not describe collective excitations. This means that the
sharp peaks in Fig. 2 do not correspond to poles of the denominator of $\chi({\bf k},\omega)$ but
are due to dynamical nesting properties of $\chi^{(0)}({\bf k},\omega)$. The energy scale
of the magnetic dispersion is in our case determined by the electron band dispersion, 
i.e., by the kinetic
energy of the electrons and not by the Heisenberg constant $J$. This is especially clear at the
point $(\pi,0)$: In the RPA $J$ drops out in $\chi$ and the energy of the magnetic exitation
is soley due to the kinetic energy of the electrons.  Using different band parametrizations 
(for instance, those given in Table I of Ref. \cite{Norman}) we found very similar 
results as in Figs. 1-2 showing that these results are generic for doped cuprates.  

$\chi''({\bf k},\omega)$ has recently been determined by
RIXS measurements along the $[1,0]$ direction for several copper oxides and 
successfully interpreted 
within a local-moment model including damping. \cite{LeTacon} Fig. 2 and its comparison with 
Ref. \cite{LeTacon} shows that  
the nature of the magnetic excitations, 
their dispersion and their band width are similar in the itinerant and in the local moment approach,
though the underlying phyics is quite different. One reason for this is that
$J$ and the effective hopping $t$ have very similar values of about 150 meV. In the $[1,1]$
direction the two approaches may show larger differences because the itinerant model predicts
in this case rather broad and irregular line shapes except for a small critical region around
$(\pi,\pi)$. 

\begin{figure}[t] 
\vspace*{-5ex}
\includegraphics[angle=270,width=6.0cm]{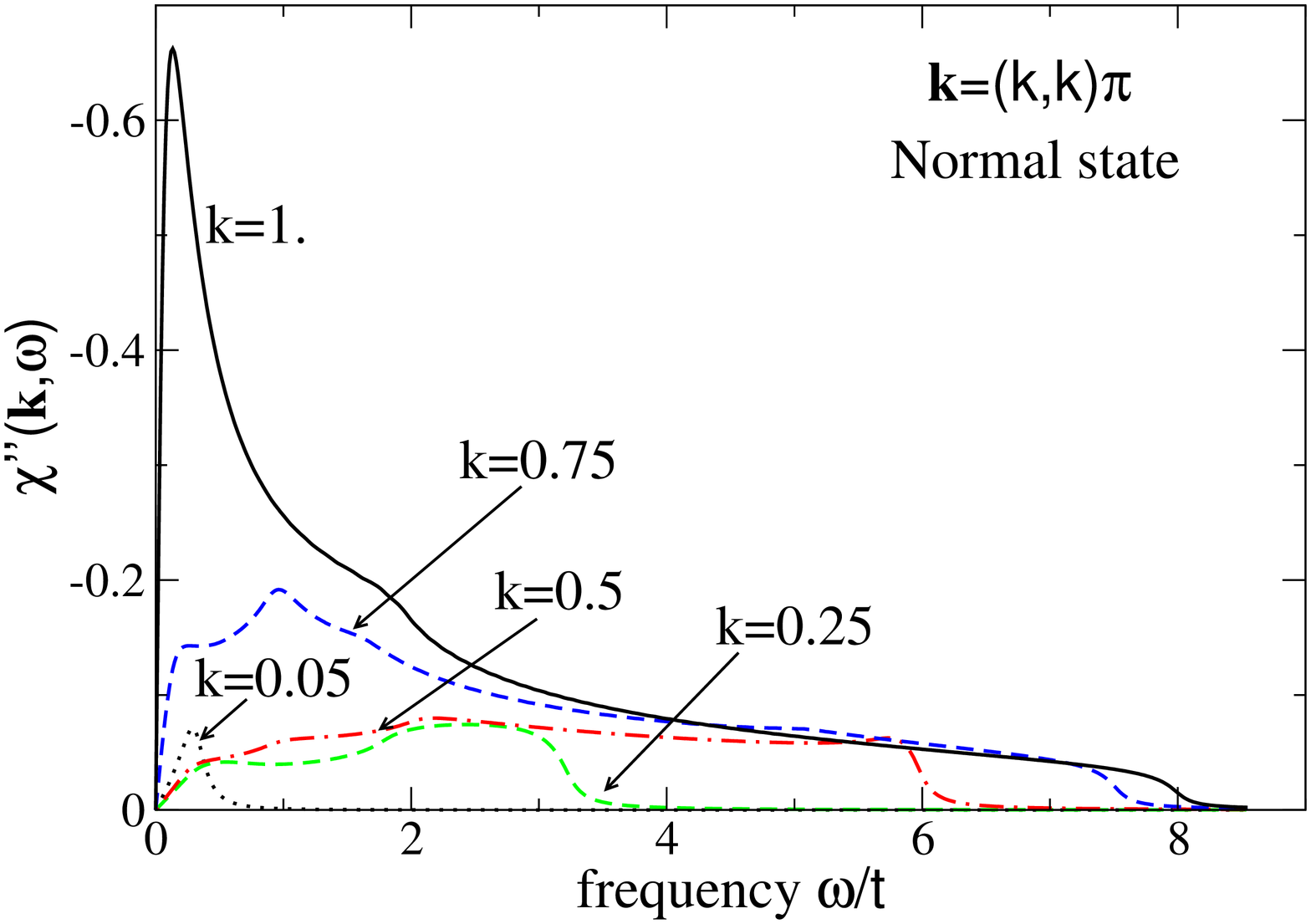}
\vspace*{-3ex}
\centerline{\includegraphics[angle=270,width=6.0cm]{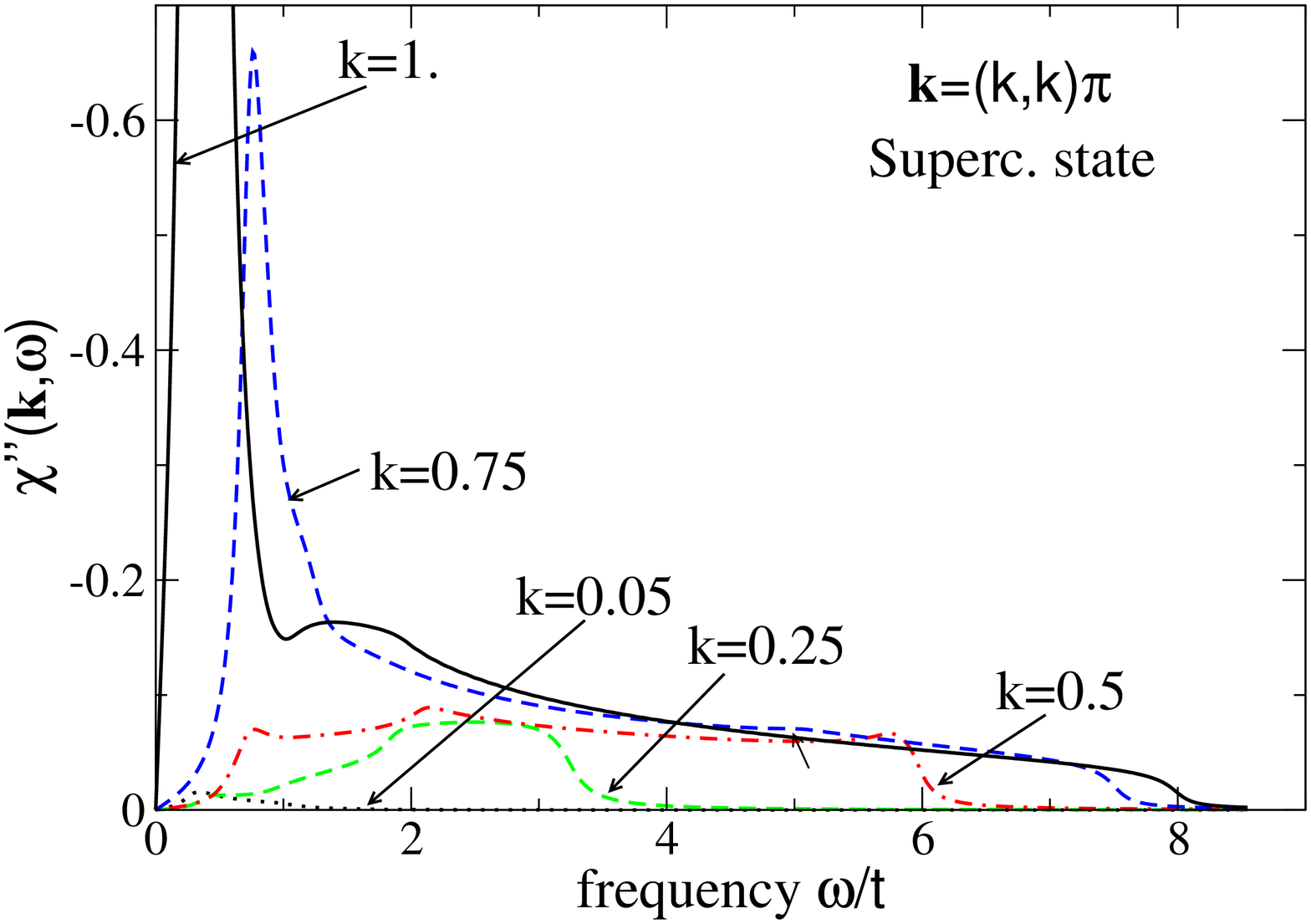}}
\vspace{-0.cm}
\caption{\label{fig:1}
(Color online)
$\chi''({\bf k},\omega)$ along the $[1,1]$ direction in the 
normal (upper diagram)
and superconducting (lower diagram) state at T=0.
}
\end{figure}

\begin{figure}[t] 
\vspace*{-2ex}
\includegraphics[angle=270,width=6.0cm]{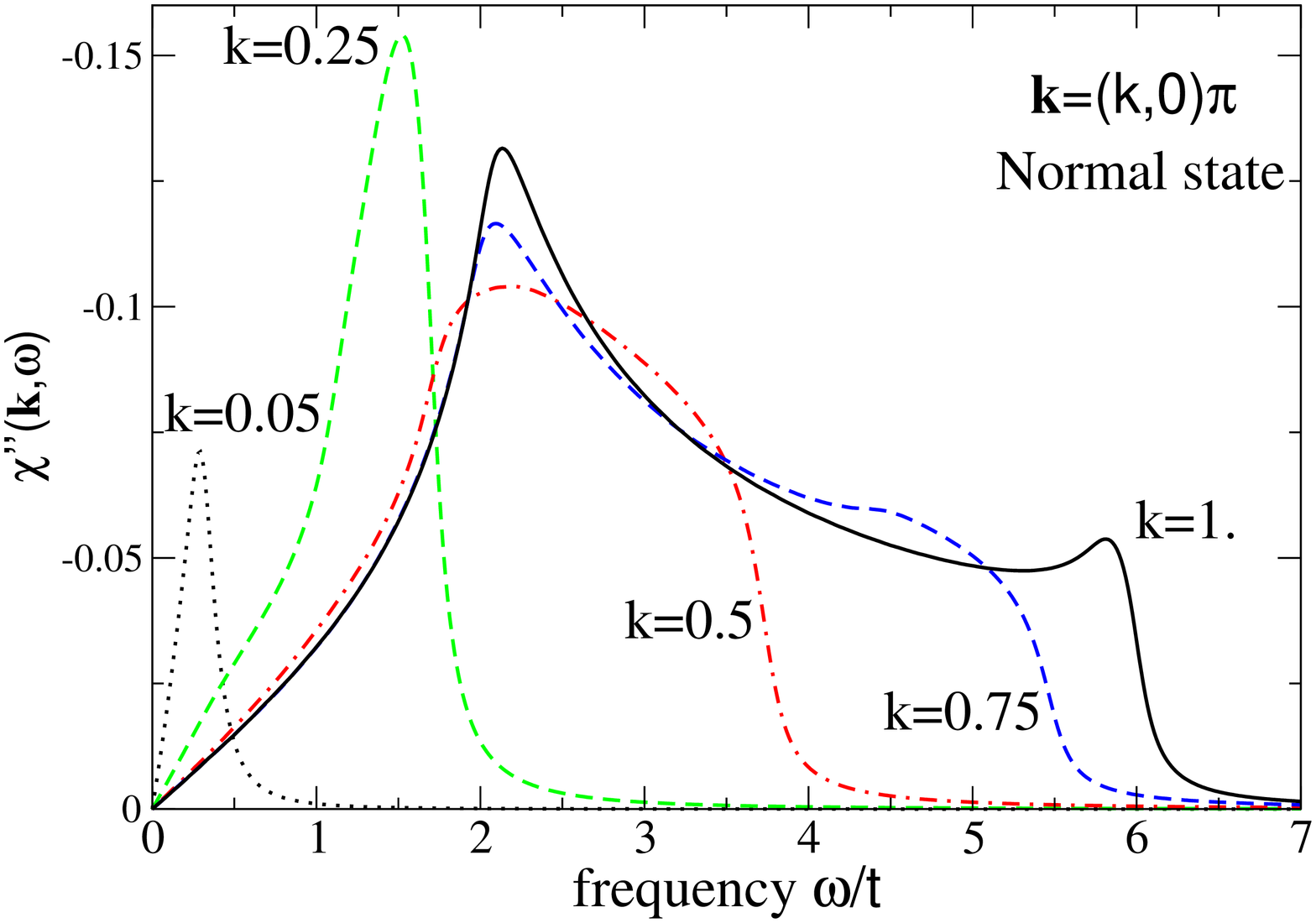}
\vspace*{-2ex}
\centerline{\includegraphics[angle=270,width=6.0cm]{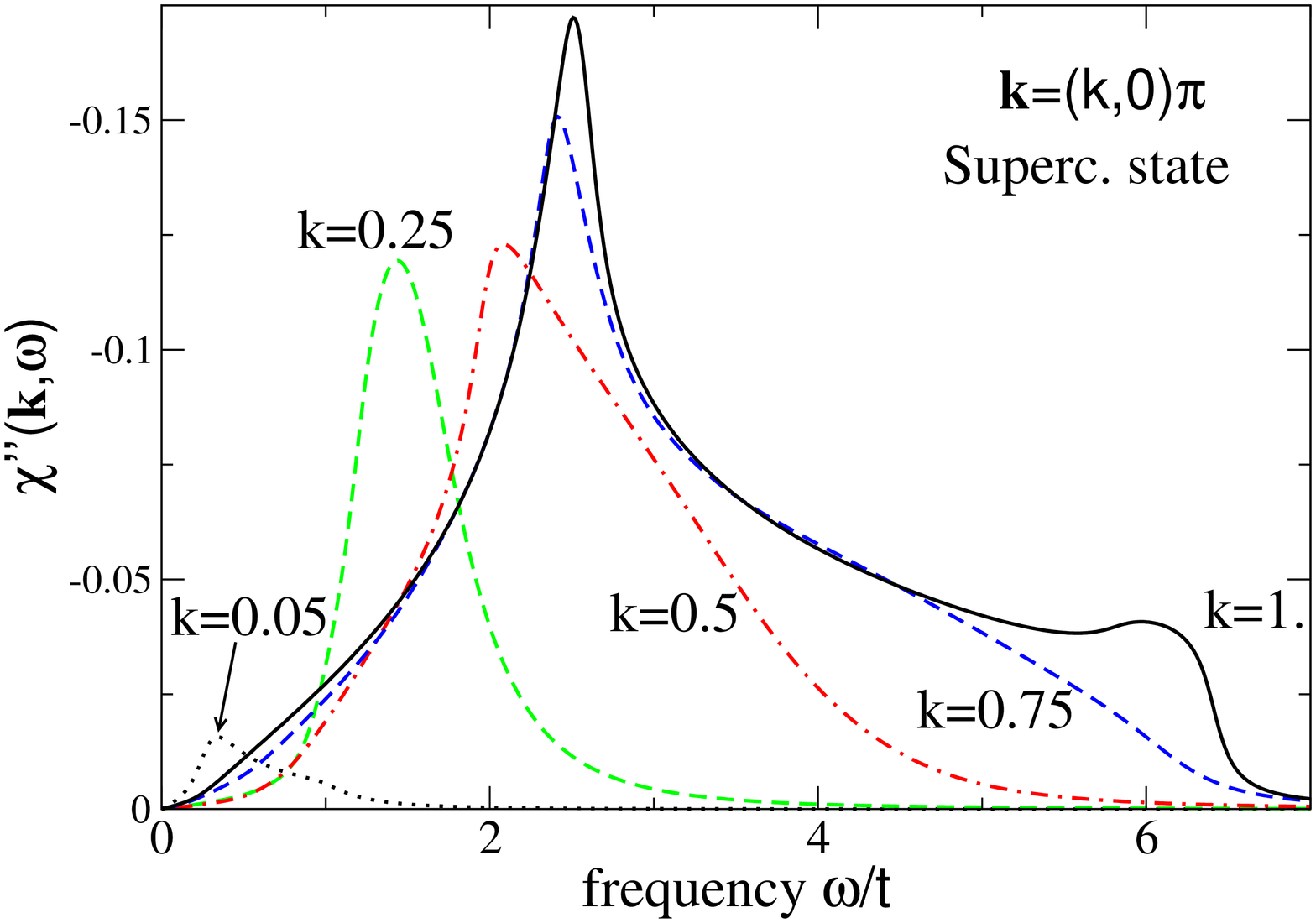}}
\vspace{-0.cm}
\caption{\label{fig:2}
(Color online)
$\chi''({\bf k},\omega)$ along the $[1,0]$ direction in the 
normal (upper diagram)
and superconducting (lower diagram) state at T=0.
}
\end{figure}

\begin{figure}[t] 
\vspace*{-5ex}
\includegraphics[angle=270,width=6.0cm]{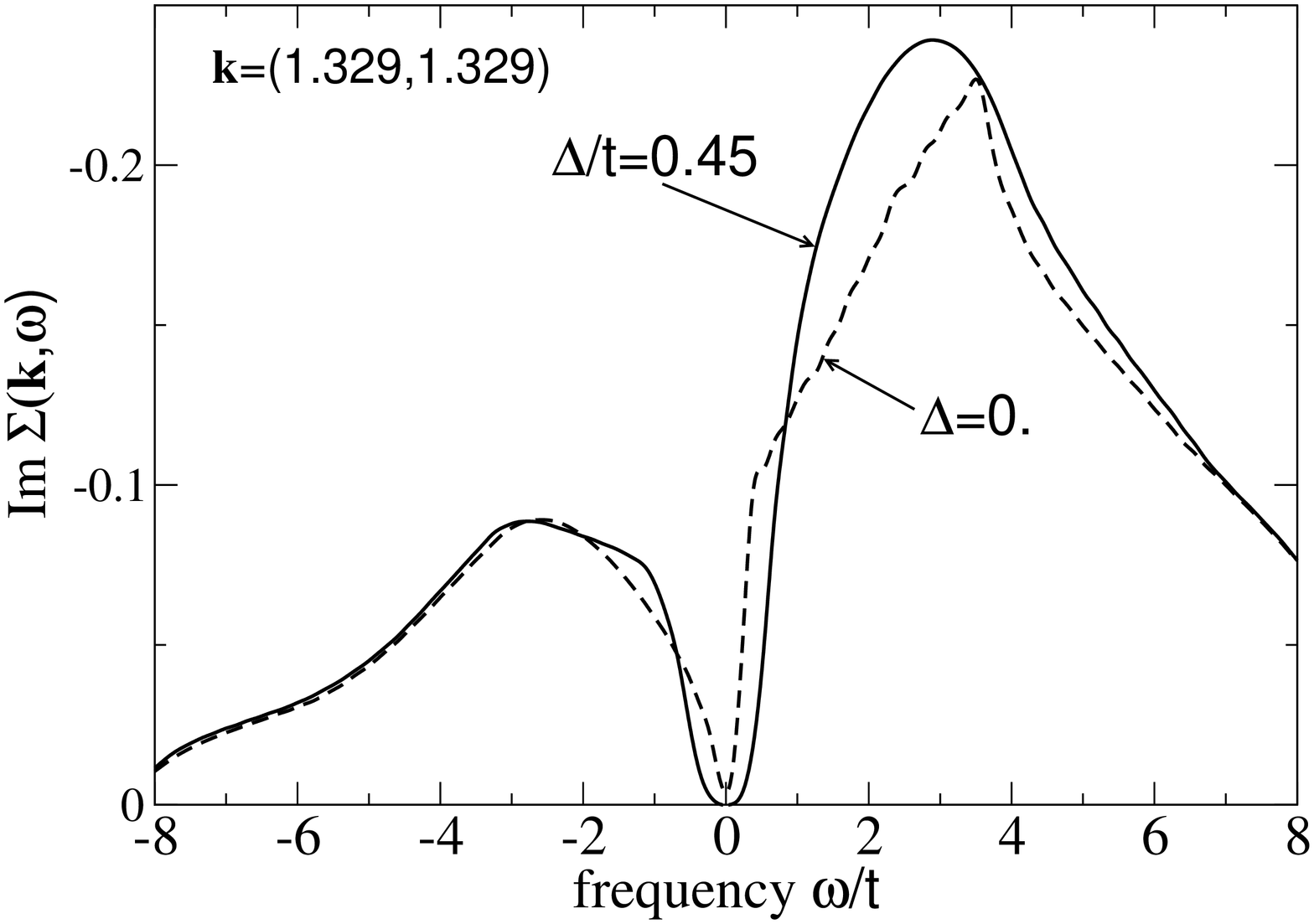}
\vspace*{-3ex}
\centerline{\includegraphics[angle=270,width=6.0cm]{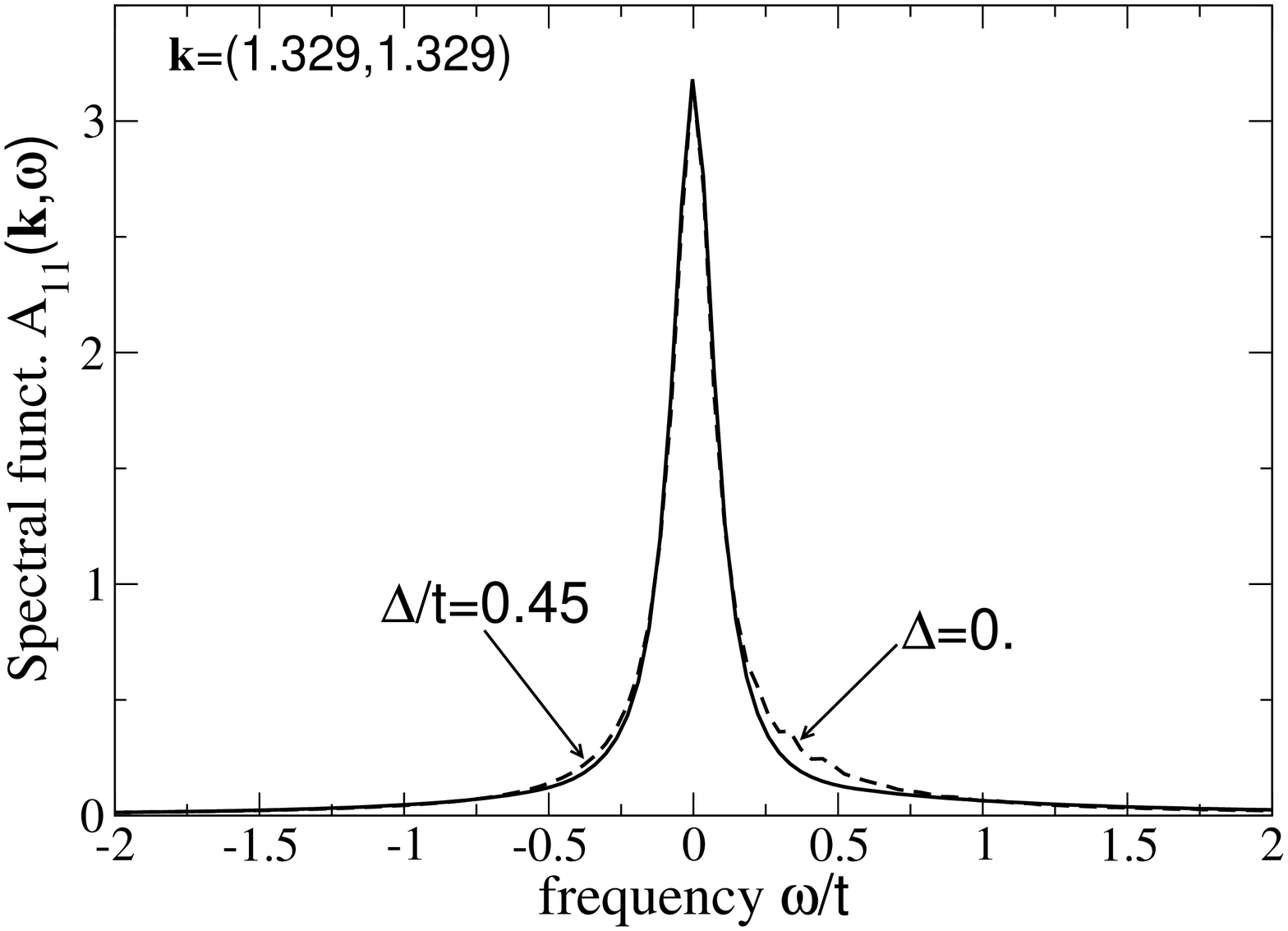}}
\vspace{0.cm}
\caption{\label{fig:3}
Imaginary part of the self-energy (upper diagram) and spectral function 
(lower diagram)
for the momentum on the Fermi line in $[1,1]$ direction in the normal and 
superconducting state.
}
\end{figure}

Fig. 3 depicts the imaginary part of $\Sigma({\bf k},\omega)$ (upper diagram) 
and
the spectral function $A_{11}({\bf k},\omega)$ (lower diagram) for the momentum 
on the Fermi
line along the [1,1] direction. In the normal state $Im \Sigma({\bf k},\omega)$
is zero at $\omega=0$, increases away from zero due to scattering of electrons
with spin fluctuations, and finally  decreases due to the finite band width.
In the superconducting state electron scattering is impeded by the gap leading
to a depletion of spectral weight around $\omega = 0$ in $Im \Sigma$ .
On the other hand superconductivity causes a large shift of spectral weight
of spin fluctuations towards low energies, especially near $(\pi,\pi)$.
Momentum transfers of about $(\pi,\pi)$ lead in general to scattered electrons
far away from the Fermi line for the considered  intial electron momentum.
The result is the superconductivity-induced peak around $\omega/t \sim 2.$, 
i.e.,
well away from the Fermi energy. The spectral function $A_{11}({\bf k},\omega)$
consists of a simple peak at $\omega =0$ without any sidebands. 

Fig. 4 shows the same as Fig. 3 but for the electron momentum
on the Fermi line along the $(\pi,0)-(\pi,\pi)$ direction. Now there are important
electronic transitions between the surroundings of $(\pi,0)$ and $(0,\pi)$ with momentum
transfers near $(\pi,\pi)$, i.e., where spin fluctuations have a large
spectral weight and low energies. In the normal state the peak
due to critial scattering near $(\pi,\pi)$ in the upper part of Fig. 1
causes the low-lying peaks in the imaginary part of
$\Sigma({\bf k},\omega)$, as seen in Fig. 4. In the superconducting state 
the same scattering processes are even much stronger due to the appearance
of the resonant mode which carries a large amount of spectral weight.
As a result the distance between the two low-lying peaks in $Im \Sigma({\bf k},\omega)$
in the normal state increases due to the superconducting gap and at the
same time the peaks gain a lot of strength. The corresponding spectral
functions reflect the low-lying peaks in form of sidebands,  
which are strongly pronounced especially in the superconducting state.
The position of the sideband peaks would be
in the absence of dispersion the 
sum of half of the superconducting gap plus the resonance frequency.
The dispersion of electrons and of the resonance peak causes 
an additional shift to the positions shown in the figure.
\begin{figure}[t] 
\vspace*{-5ex}
\includegraphics[angle=270,width=6.0cm]{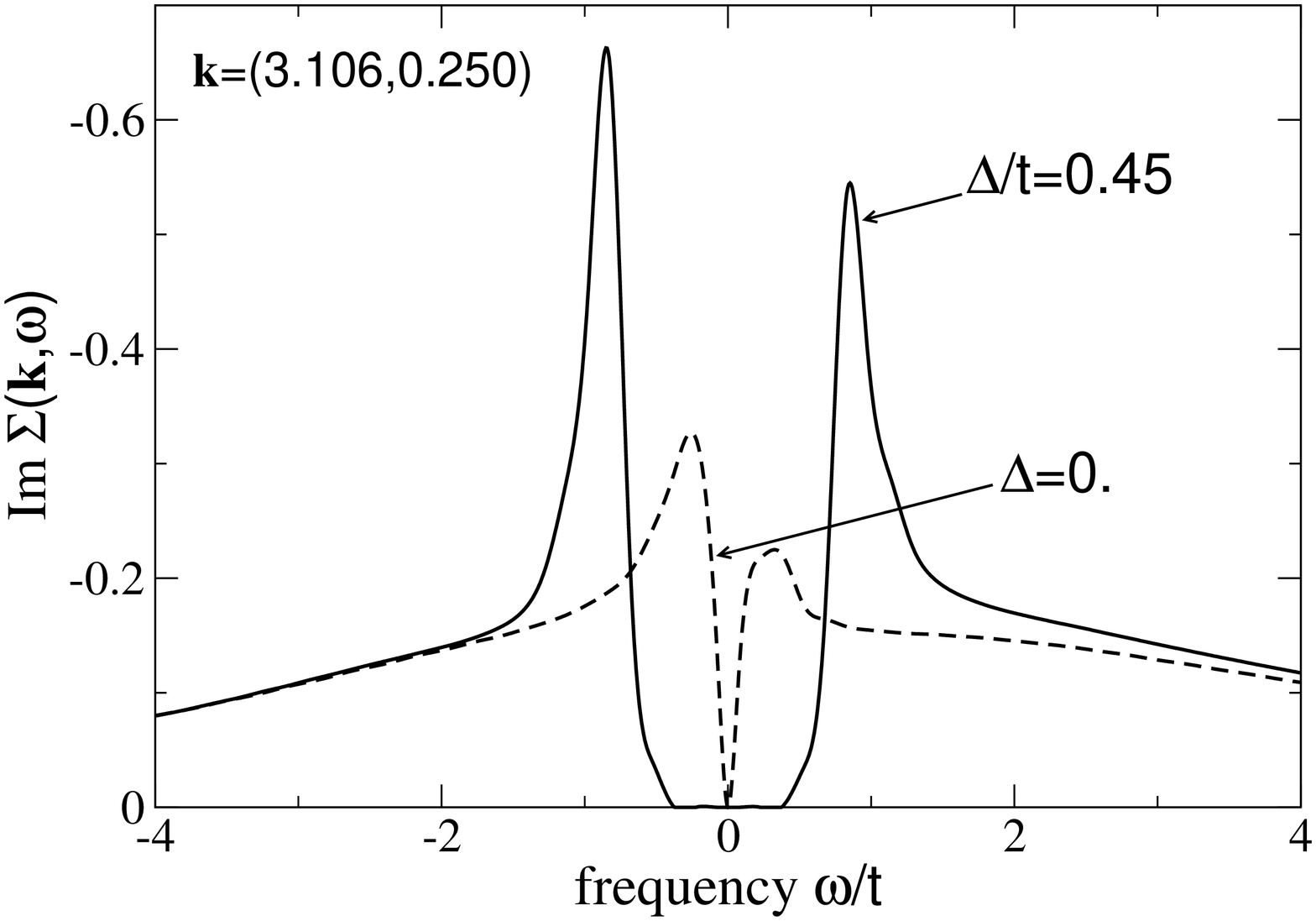}
\vspace*{-3ex}
\centerline{\includegraphics[angle=270,width=6.0cm]{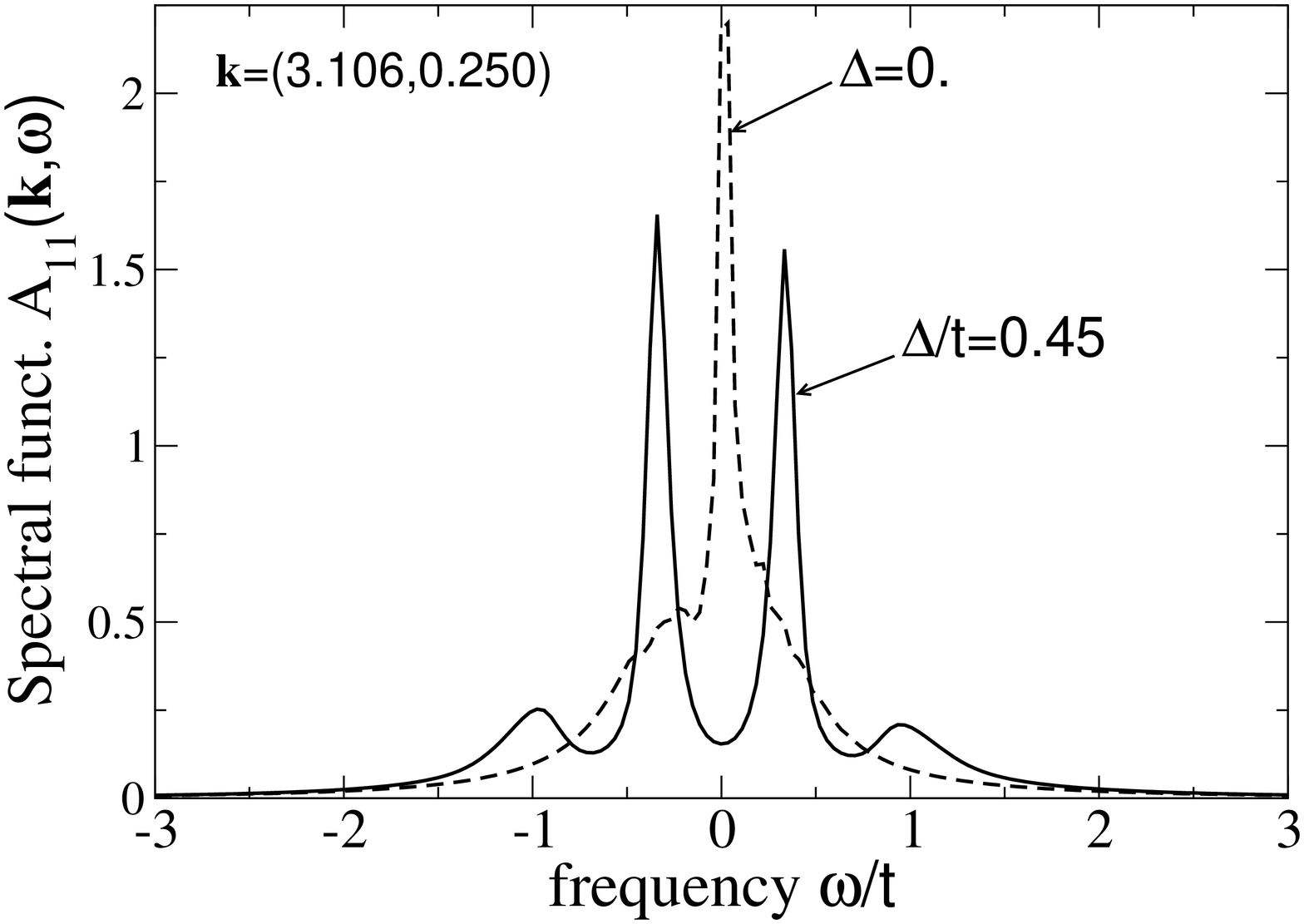}}
\vspace{0cm}
\caption{\label{fig:4}
Imaginary part of the self-energy (upper diagram) and spectral function 
(lower diagram)
for the momentum on the Fermi line in $(\pi,0)-(\pi,\pi)$ direction in the normal and 
superconducting state.
}
\end{figure}
\begin{figure}[t] 
\includegraphics[angle=270,width=9.0cm]{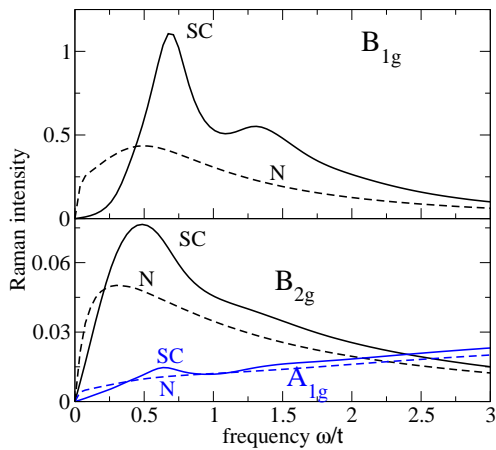}
\vspace*{-5ex}
\vspace{0.4cm}
\caption{\label{fig:5}
(Color online)
Raman spectra with B$_{1g}$ (upper part), B$_{2g}$ and A$_{1g}$ symmetries 
(lower part) in the normal (dashed lines) and superconducting 
(solid lines) state.  
}
\end{figure}

Finally, Fig. 5 gives our results for the Raman intensity $-Im \chi_\alpha(\omega+i\eta)$
where $\alpha$ denotes one of the three symmetries B$_{1g}$, B$_{2g}$ and A$_{1g}$.
In each case the dashed line corresponds to the normal
and the solid line to the superconducting state. Let us first discuss the B$_{1g}$ spectrum
(upper part in Fig. 5).
In the free case, defined by $\Sigma= 0$,
the Raman intensity would consist in the normal state of a $\delta$-function at $\omega = 0$ 
(not shown in the figure) and a Lorentzian-like pair breaking peak at the unrenormalized energy $2\Delta$. 
Turning on $\Sigma({\bf k},\omega)$ the $\delta$-function is replaced by a continuum  
in the normal state which increases rather abruptly from
zero at zero frequency to a finite value and then decreases very slowly towards higher
frequencies. This is the result of electron scattering from spin fluctuations.
The solid curve, referring to the superconducting state and B$_{1g}$ symmetry, exhibits two maxima. 
The first one is due to the renormalized pair
breaking peak, the second one due to the additional emission of a magnetic excitation, for instance,
the resonance mode. The position of the additional high-frequency peak can also be estimated from
the spectral function A$_{11}$. According to Eq.(\ref{chi}) the Raman intensity at frequency $\omega$
is essentially be obtained 
as a folding of two spectal functions displaced by the frequency $\omega$. As a result
the pair breaking peak arises from tansitions between the two dominating peaks in Fig. 4. A smaller 
peak is expected from transitions between one of the large peaks and a sideband peak yielding
a frequency of about 1.3 in good agreement with Fig. 5.   
The ratio of the positions of the two peaks is about 1.9 
which is near the experimental value. \cite{Li} 
 We stress that the position
and the intensity of the second peak in this spectrum is not fitted but a consequence of the
parameters $J$ and $\Delta$ determined such that the observed resonance frequency and the pair breaking peak
are at least approximately reproduced. Using the electronic parameters of Ref. \cite{Zeyher2} the
additional peak in the B$_{1g}$ spectrum is practically absent. The reason for this is that the
self-energy in this case is substantially smaller than in our case. The difference in the magnitude of the 
self-energies can at least partially be traced back to the large difference in the resonance frequencies,
namely 56 meV in HgBa$_2$CuO$_{4+\delta}$ and 40 meV in YBCO for nearly optimally doped samples.

The B$_{2g}$ spectrum (contained in the lower diagram in Fig. 5) is in the normal state similar
to that of the B$_{1g}$ symmetry except that its intensity is smaller by one order of magnitude. In the 
superconducting state it shows a pair breaking peak but no additional peak at higher frequencies,
in agreement with experiment. \cite{Li2}
The reason for this can easily be understood:
The form factor in the B$_{2g}$ symmetry weights electrons near the nodal direction strongly.
There the gap and also the self-energy are small. As a result the pair breaking moves somewhat to lower
frequencies compared to the B$_{1g}$ symmetry and no additional peak appears as a result of self-energy effects.
For completeness we also show the A$_{1g}$ spectrum (lower diagram in Fig. 5) where Coulomb screening has 
been taken into account. Here we encounter the well-known and hitherto unsolved problem that  
its intensity is too small by about one order of magnitude compared
to experiment. \cite{Gallais} Thus we will exclude it from our discussion.

We used in our treatment a broken Cooper pair times a spin fluctuation as final states. Using the same interactions
one may also consider a competing process where two spin fluctuations or paramagnons appear 
as final states. \cite{Li} For zero or very small doping such a localized description is certainly appropriate
whereas in the  optimal or overdoped regime a description in terms of itinerant electrons should be a better 
choice.
For instance, it allows easily to understand why the additional high-energy peak shows a strong dependence
on $T_c$ similar like the resonance mode \cite{Li} or why its existence is tied to the existence of the 
pair breaking peak
and the depletion region below it. It also has been argued \cite{Venturini} that two-magnon processes are
neglegible small in the B$_{1g}$ and B$_{2g}$ configurations and thus may be discarded in these cases. 

In summary, we have shown that an itinerant picture of magnetism
may account for the observed dispersive magnetic excitations in $[1,0]$ direction in doped cuprates
and also for the observed additional structure above the pair breaking peak in electronic Raman
scattering in HgBa$_2$CuO$_{4+\delta}$. 

The authors thank M. Le Tacon for a critical reading of the manuscript.
They also acknowledge useful discussions with Y. Li, M. Le Tacon, A. Schnyder
and B. Keimer.
R.Z. and A.G. are grateful to the Dep. de F\'{\i}sica (Rosario) and
the MPI-FKF (Stuttgart), respectively, for hospitality and financial support.


\end{document}